\documentstyle[pre,aps,multicol]{revtex}
\input{epsf.tex}

\newcommand{\mpictf}[4]{
\begin{figure}[H]
\setlength{\epsfxsize}{#3}
\centerline{\mbox{\epsffile{#2}}}
\vspace{2mm}
\protect\caption{ \protect\label{#1} #4}
\end{figure} }

\title{Multi-component optical solitary waves}
\author{Yuri S. Kivshar$^1$, Andrey A. Sukhorukov$^1$, 
Elena A. Ostrovskaya$^1$, Tristram J. Alexander$^1$,
Ole Bang$^{1,2}$,\\ 
Solomon M. Saltiel$^3$,
Carl Balslev Clausen$^2$, and Peter L. Christiansen$^2$}

\address{
$^1$ Optical Sciences Centre, Australian National University, 
     Canberra ACT 0200, Australia \\
$^2$ Department of Mathematical Modeling, Technical University of Denmark,
Lyngby DK-2800, Denmark\\
$^3$ Quantum Electronics Department, Faculty of Physics, University of
Sofia, Sofia 1164, Bulgaria
}

\begin{document} 
\maketitle 
\begin{abstract}
We discuss several novel types of multi-component (temporal and spatial)
envelope solitary waves that appear in fiber and waveguide 
nonlinear optics.  In particular, we describe multi-channel solitary waves 
in bit-parallel-wavelength fiber transmission systems for high performance 
computer networks, multi-colour parametric spatial solitary waves due to 
cascaded nonlinearities of quadratic materials, and quasiperiodic envelope 
solitons due to quasi-phase-matching in Fibonacci optical superlattices.
\end{abstract}
\begin{multicols}{2}
\narrowtext
\section{Introduction}

Rapid progress in the design and manufacture of optical fiber 
systems is a result of worldwide 
demand for ultra-high bit-rate optical communications. This explains 
the growing interest of the soliton community in soliton-based optical 
fiber communication systems. This area of research was considerably 
advanced in recent years~\cite{hasegawa_book}. The most remarkable results
include the application of the concept of the dispersion management to 
{\em temporal optical solitons} and
soliton-based optical transmission systems, and the discovery of the 
so-called {\em dispersion managed soliton}. 
High-speed optical communications require effective components such as
high-performance broadband computer networks that can be developed by
employing the concept of the bit-parallel-wavelength (BPW) pulse transmission 
that offers many of the advantages of both parallel 
fiber ribbon cable and conventional wavelength-division-multiplexing 
(WDM) systems~\cite{review}. 

Expanding development in the study of the soliton fiber systems has been observed 
in parallel with impressive research on their spatial counterparts, 
optical self-trapped beams or {\em spatial optical solitons}. 
One of the key concepts in this field came from 
the theory of multi-frequency wave mixing and cascaded 
nonlinearities where a nonlinear phase shift is produced as a result of the 
parametric wave interaction~\cite{stegeman,chi2_review}. 
In all such systems, the nonlinear interaction between the waves 
of two (or more) frequencies is the major physical effect that can 
support coupled-mode multi-frequency solitary waves.

The examples of temporal and spatial solitons mentioned above have
one common feature: they involve the study of solitary waves in
multi-component nonlinear models.
The main purpose of this paper is to overview several different 
physical examples of multi-mode and/or multi-frequency solitary waves 
that occur for the pulse or beam propagation
in nonlinear optical fibers and waveguides.  
For these purposes, we select three different cases:
multi-wavelength solitary waves 
in bit-parallel-wavelength optical fiber links, 
multi-colour spatial solitons due to multistep cascading in optical 
waveguides with quadratic nonlinearities, and quasiperiodic solitons 
in the Fibonacci superlattices. We believe these examples display both 
the diversity and richness of the multi-mode soliton systems, and they will 
allow further progress to be made in the study of nonlinear waves in 
multi-component nonintegrable physical models.

\section{Temporal and spatial solitons}

Because the phenomenon of the long-distance propagation of {\em temporal optical
solitons} in optical fibers~\cite{hasegawa_book} is known to 
a much broader community of researchers in optics and nonlinear physics, 
first we emphasize {\em the difference between temporal and spatial solitary 
waves}. Indeed, for a long time stationary beam propagation
in planar waveguides has been considered somewhat similar to the pulse
propagation in fibers.  This approach is based on the so-called
{\em spatio-temporal analogy} in wave propagation, meaning that the
propagation coordinate $z$ is treated as the evolution variable and the
spatial beam profile along the transverse direction in 
waveguides, is similar to the temporal pulse profile in 
fibers.  This analogy is based on
a simple notion that both beam evolution and pulse propagation can be
described by the cubic nonlinear Schr\"odinger (NLS) equation.
  
However, contrary to the widely accepted opinion, there
is a crucial difference between temporal and spatial solitons.  
Indeed, in the case of the nonstationary
pulse propagation in fibers, the operation wavelength is usually selected
near the zero point of the group-velocity dispersion. This means that the
absolute value of the fiber dispersion is small enough to be compensated by
a weak nonlinearity such as that produced by the (very weak) Kerr effect in
optical fibers which leads to a relative nonlinearity-induced change in the
refractive index. Therefore, nonlinearity in
such systems  is always weak and it should be well modeled by a cubic NLS equation which is known to be integrable by means of the
inverse-scattering technique. 
However, for very short (e.g., fs) pulses the cubic
NLS equation describing the long-distance propagation of pulses should be
corrected to include additional terms that would account for such effects as higher-order
dispersion, Raman scattering, etc. All such corrections can be taken into
account with the help of the perturbation theory~\cite{hasegawa_book}. 
Thus, in fibers nonlinear effects are weak and they become important only 
when dispersion is small (near the zero-dispersion point) affecting the 
pulse propagation over large distances (of order of hundreds of meters 
or even kilometers).  
The situation changes dramatically when we consider the propagation 
of multi-wavelength pulses with almost equal group velocities.  
The corresponding model is described by a nonintegrable and rather
complicated system of coupled NLS equations, 
which we briefly discuss below.

In contrary to the pulse propagation in optical fibers, the physics underlying
the stationary beam propagation in planar waveguides and bulk media is
different. In this case an optical beam is generated by a continuous wave (CW)  source
and it is time independent. However, when the beam evolves with the
propagation distance $z$, it diffracts in the transverse spatial directions.
Then, a nonlinear change in the refractive index should
compensate for the beam spreading caused by diffraction {\em which is not a
small effect}. That is why to observe spatial solitons as
self-trapped optical beams, much larger nonlinearities 
are usually required, and very often such nonlinearities 
are not of the Kerr type (e.g. they saturate at higher intensities). 
This leads to the models of generalized nonlinearities with
the properties of solitary waves different from those described by the
integrable cubic NLS equation. Propagation distances involved in
the phenomenon of the beam self-focusing and spatial soliton propagation are of
the order of millimeters or centimeters.
To achieve such large nonlinearities, one needs to use the optical materials
with large nonlinearity-induced refractive index. One of the possible way
to overcome this difficulty is to use the so-called 
{\em cascaded nonlinearities} of noncentrosymmetric 
optical materials where nonlinear effects are accumulated due to 
parametric wave interaction under the condition of the wave phase matching.  
Such parametric wave-mixing effects generate novel classes of spatial
optical solitons where resonant  parametric coupling between the
envelopes of two (or more)  beams
of different frequencies  supports  stable spatially localised waves even in
a bulk medium (see details in Ref.~\cite{stegeman}).  
It is this kind of multi-component solitary waves that we discuss below.

\section{BIT-PARALLEL-WAVELENGTH SOLITONS}

A growing demand for high-speed computer communications requires an
effective and inexpensive computer interconnection. One attractive alternative to the conventional WDM systems is BPW  
 single-mode fiber optics links for very high bandwidth computer 
communications~\cite{review}. They differ from the 
 WDM schemes in that no parallel to serial conversion is 
necessary, and parallel pulses are launched simultaneously on different 
wavelengths.

When the pulses of different wavelengths are transmitted simultaneously,
the cross-phase modulation can produce an interesting 
{\em pulse shepherding effect}~\cite{yeh1}, 
when a strong ("shepherd") pulse enables the
manipulation and control of pulses co-propagating on different wavelengths 
in a multi-channel optical fiber link.  

To describe the simultaneous transmission of $N$ different wavelengths 
in a nonlinear optical fiber, we follow the standard derivation~\cite{agrawal}
and obtain a system of $N$ coupled nonlinear Schr\"odinger
(NLS) equations $(0 \leq j \leq N-1$): 
\begin{equation} \label{eq:NLS_dim}
 \begin{array}{l}
  {\displaystyle 
   i \frac{\partial A_j}{\partial z} 
   + i v_{gj}^{-1} \frac{\partial A_j}{\partial t} 
   - \frac{\beta_{2j}}{2} \frac{\partial^2 A_j}{\partial t^2} 
  } \\*[12pt] {\displaystyle \qquad
    + \chi_j \left( |A_j|^2 + 2 \sum_{m \neq j} |A_m|^2 \right) A_j
    = 0,
  } \end{array}
\end{equation} 
where, for the $j-$th wave, $A_j(z,t)$ is the
slowly varying envelope, $v_{gj}$ and $\beta_{2j}$  are the group velocity
and group-velocity dispersion,  respectively, and the nonlinear
coefficients  $\chi_j$ characterize the Kerr effect.  
Equations~(\ref{eq:NLS_dim})
do not include the fiber loss, since the fiber lengths involved in
bit-parallel links are only a small fraction of the attenuation length.

We measure the variables in the units of the central wavelength channel
(say, $j=0$), 
and obtain the following normalized system of the $N$ coupled NLS
equations,
\begin{equation} \label{eq:NLS}
 \begin{array}{l}
  {\displaystyle  
    i \frac{\partial u_j}{\partial z}  
    + \frac{1}{2} \alpha_j \frac{\partial^2 u_j}{\partial t^2} 
  } \\*[12pt] {\displaystyle \qquad
    + \gamma_j \left(|u_j|^2 + 2 \sum_{m\neq j} |u_m|^2\right) u_j
    = 0,
  } \end{array}
\end{equation}
where $u_j = A_j/\sqrt{P_0}$, $P_0$ is the incident optical power in the
central channel, $\alpha_j \equiv (\beta_{2j}/|\beta_{20}|)$, 
$\gamma_j \equiv \chi_j/\chi_0$, so that $\alpha_0 = \gamma_0 =1$.  For the
operating wavelengths spaced $1\,$nm apart within the band $1530 \div 1560$ nm, the
coefficients $\alpha_j$ and $\gamma_j$ are different but close to $1$. Initially, in Eq.~(\ref{eq:NLS}), we omit the mode walk-off effect described
by the parameters $\delta_j = v_{g0}^{-1} - v_{gj}^{-1}$ (so that $\delta_0=0$). This effect will be analysed later in this section.

To analyze the nonlinear modes, i.e. localized states of the BPW model
(\ref{eq:NLS}), we look for stationary solutions in the form,
\begin{equation} \label{eq:station}
  u_j(z,t) = u_j(t) e^{i\beta_j z},
\end{equation}
and therefore obtain the system of equations for the normalized mode
amplitudes,
\begin{equation} \label{eq:NLS_Nn}
 \begin{array}{r}
 { \displaystyle
   \frac{1}{2} \frac{d^2 u_0}{dt^2} + \left( |u_0|^2 +
   2\sum_{n=1}^{N-1}|u_n|^2 \right) u_0
   = \frac{1}{2} u_0,
 } \\*[15pt] { \displaystyle
   \frac{1}{2} \alpha_n \frac{d^2u_n}{dt^2} 
   + \gamma_n \left( |u_n|^2 
   + 2\sum_{m \neq n} |u_m|^2 \right) u_n 
   = \lambda_n u_n,
 } \end{array}
\end{equation}
where $n =1,2, \dots, N-1$, the amplitudes and time are measured in the
units of $\sqrt{2\beta_0}$ and $(2\beta_0)^{-1/2}$, respectively, and
$\lambda_n = \beta_n/2\beta_0$.

System (\ref{eq:NLS_Nn}) has {\em exact analytical solutions} for $N$ coupled
components, the so-called {\em BPW solitons}. Indeed, looking for solutions
in the form $u_0(t) = U_0 \; {\rm sech} \, t$, 
$u_n(t) = U_n \;\; {\rm sech} \; t$, 
we obtain the constraint $\lambda_n = \alpha_n/2$, and a system
of $N$ coupled algebraic equations for the wave amplitudes,
\[
   U_0^2 + 2 \sum_{n=1}^{N-1} U_n^2 = 1,   \;\; 
   U_n^2 + 2 \sum_{m \neq n} U_m^2 = \alpha_n/\gamma_n.
\]
In a special symmetric case, we take $\alpha_n = \gamma_n =1$,  
and the solution of those equations is
simple~\cite{yeh5}:\\
 $U_0 = U_n \equiv U_* = [1 + 2(N-1)]^{-1/2}$.

Analytical solutions can also be obtained in the {\em linear limit}, 
when the central frequency pulse (at $n=0$) is large.
Then, linearizing Eqs.~(\ref{eq:NLS_Nn}) for small $|u_n| \ll |u_0|$, we
obtain a decoupled nonlinear equation for $u_0$ and $N-1$ 
decoupled linear equations for $u_n$.  Each of the latter possess a
localized solution provided $\lambda_n = \lambda_n^{(0)}$, where
$\lambda^{(0)}_n = (\alpha_n/8) [1-\sqrt{1+16(\gamma_n/\alpha_n)}]^2$. 
In this limit the central soliton pulse $u_0$ (``shepherd pulse'') can be 
considered as inducing an effective waveguide that supports 
a fundamental mode $u_n$ with the corresponding cutoff $\lambda^{(0)}_n$. 
Since, by definition, the parameters $\alpha_n$ and $\gamma_n$ are close 
to $1$, we can verify that the soliton-induced waveguide 
supports maximum of two modes (fundamental and the first excited one).
This is an important physical result that explains the effective
robustness of the pulse guidance by the shepherding pulse.

\mpictf{fig:bpw1}{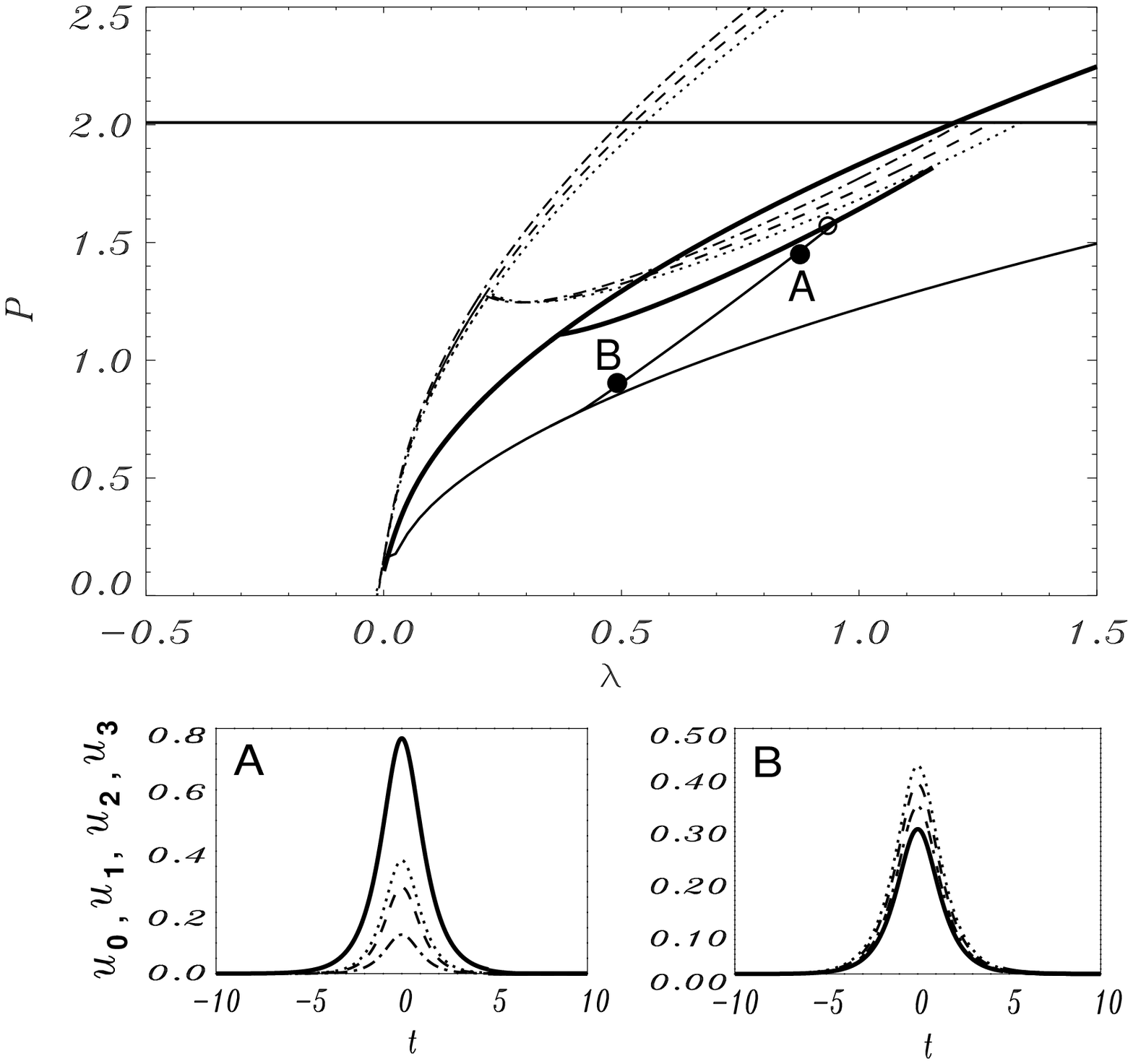}{8.0cm}{
Bifurcation diagram (top) and two examples of the 
stationary BPW soliton solutions for $N=4$.}

To  demonstrate a number of unique properties of the multi-channel 
BPW solitons, we consider the case $N=4$ in more details. A comprehensive 
discussion of the case $N=2$ can be found in the preprint~\cite{lena_lanl}.
We select the
following set of the normalized parameters: 
$\alpha_{0,3} = \gamma_{0,3} = 1$, 
$\alpha_1=\gamma_1=1.1$, and $\alpha_2 =\gamma_2=1.05$.  
Solitary waves of this four-wavelength BPW system can be found numerically as 
localized solutions of Eqs.~(\ref{eq:NLS_Nn}).  
Figure~\ref{fig:bpw1} presents the lowest-order families of such
localized solutions. In general, they are characterized by $N-1$ parameters, 
but we can capture the characteristic features by presenting power
dependencies along the line $\lambda \equiv \lambda_1=\lambda_2=\ldots$ 
in the parameter space $\{\lambda_n\}$. The power of the central-wavelength
component ($n=0$) does not depend on $\lambda$ (straight line $P_0=2$).
Thin dashed, dotted, and dash-dotted curves correspond to the 
three separate single-mode solitons of the multi-channel BPW system, (1),  
(2), and (3), respectively, shown with the corresponding branches of (0+1),  
(0+2), and (0+3) two-mode solitons. The latter curves start off from 
the bifurcation points on the $u_0$ branch at $\lambda^{(0)}_1$, 
$\lambda^{(0)}_2$, and $\lambda^{(0)}_3$, respectively. 
Close separation of those curves is the result of closeness of 
the parameters $\alpha_n$ and $\gamma_n$ for $n=1,2,3$.

Thick solid curves in Fig.~\ref{fig:bpw1} 
correspond to the two- (1+2) and three-mode (0+1+2) localized solutions. 
The latter  solutions bifurcate and give birth to four-wavelength 
solitons (0+1+2+3) (branch A-B).  Two examples of such four-wave composite 
solitons are shown in Fig.~\ref{fig:bpw1} (bottom row). The
solution B is close to an exact sech-type solution at $\lambda=0.5$ 
(described above) for $N=4$, whereas the solution A is close to that 
approximately described in the linear limit in the vicinity of 
a bifurcation point. Importantly, for different values of the parameters 
$(\alpha_n,\gamma_n)$, the uppermost bifurcation point for this branch 
(open circle in Fig. \ref{fig:bpw1}) is not predicted by a simple linear 
theory and, due to the nonlinear mode coupling, it gets shifted from 
the branch of the central-wavelength soliton (straight line) 
to a two-mode branch (0+1+2) (thick solid curve).

As a result, if we start on the right end of the horizontal branch and 
follow the lowest branches of the total power $P(\lambda)$ in 
Fig.~\ref{fig:bpw1}, we pass the following sequence of the soliton 
families and bifurcation points: 
$(0) \rightarrow (0+1) \rightarrow (0+1+2) \rightarrow
(0+1+2+3) \rightarrow (1+2+3) \rightarrow (2+3) \rightarrow (3)$. 
If the modal parameters are selected closer to each other, the first 
two links of {\em the bifurcation cascade} disappear (i.e. the 
corresponding bifurcation points merge), and the four-mode soliton 
bifurcates directly from the central-wavelength pulse, as predicted 
by the linear theory. 
Note however that the sequence and location of the bifurcation points
is a function of the cross-section of the parameter space 
$\{\lambda_n\}$, and the results presented above correspond to 
the choice $\lambda=\lambda_1=\lambda_2=\ldots$.

The qualitative picture of the cascading bifurcations preserves for other 
values of $N$. In particular, near the bifurcation point a mixed-mode 
soliton corresponds 
to the localized modes guided by the central-wavelength soliton (shepherd) 
pulse. The existence of such soliton solutions is a key concept of BPW 
transmission when the data are launched in parallel carrying a desirable 
set of bits of information, all guided by the shepherd pulse  at a selected 
wavelength.

Effects of the walk-off on the multi-channel BPW solitons seems to 
be most dangerous for the pulse alignment in the parallel links.  
For nearly equal soliton components, it was shown long time ago 
\cite{menyuk} that nonlinearity can provide an effective trapping 
mechanism to keep the pulses together. For the shepherding effect, 
the corresponding numerical simulations are presented in
Figs.~\ref{fig:bpw2}(a-d) for the four-channel BPW system. 
Initially, we launch a composite four-mode soliton as an unperturbed 
solution A [see Fig.~\ref{fig:bpw1}] of Eqs.~(\ref{eq:NLS}), 
without walk-off and centered at $t=0$. When this solution evolves 
along the propagation direction $z$ in the presence of small
to moderate relative walk-off ($\delta_n \neq 0$ for $n\neq 0$), 
its components remain strongly localized and mutually trapped 
[Fig.~\ref{fig:bpw2}(a,b)], whereas it loses some energy into radiation 
for much larger values of the relative mode walk-off 
[Fig.~\ref{fig:bpw2}(c,d)].

\mpictf{fig:bpw2}{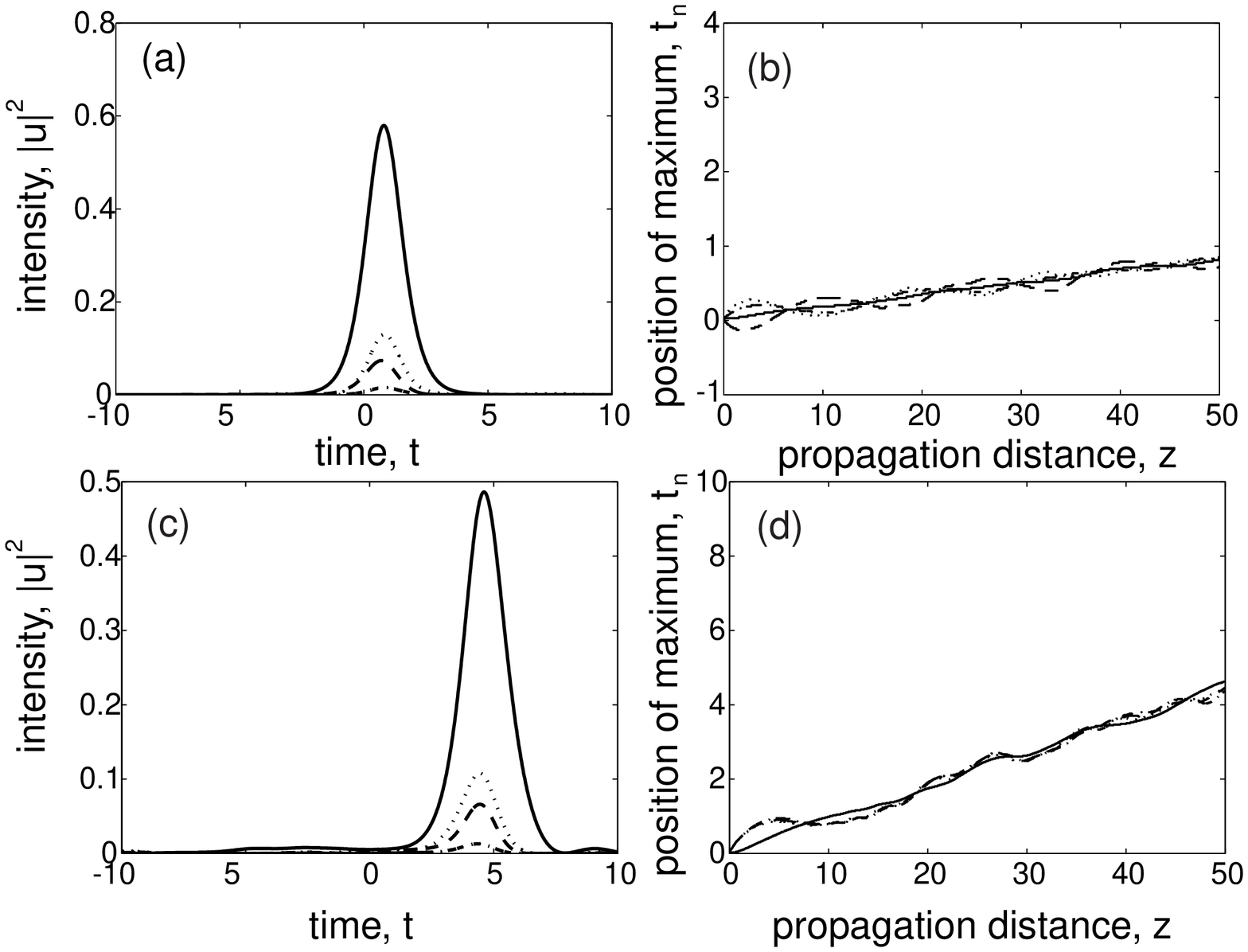}{8.0cm}{
Influence of the mode walk-off on the shepherding effect for the input 
pulse A of Fig.~\ref{fig:bpw1}. 
Shown are: output pulse at the propagation distance 
$z=50$ for (a)~$\delta_1=0.45, \delta_2=-0.35, \delta_3=0.25$, and 
(c)~$\delta_1=\delta_2=\delta_3=0.9$; 
(b,d)~--- evolution of the position of maximum for all pulse constituents. 
Solid line~--- maxima of the shepherd pulse.}

\section{PARAMETRIC OPTICAL SOLITONS DUE TO MULTISTEP CASCADING}
\subsection{Concept of multistep cascading}

Recent progress in the study of cascading effects in optical materials with
quadratic (second-order or $\chi^{(2)}$) nonlinear response has offered 
new opportunities for all-optical processing, optical
communications, and optical solitons~\cite{stegeman,chi2_review}. Most of
the studies of cascading effects employ parametric wave mixing
processes with a single phase-matching and, as a result, two-step
cascading~\cite{stegeman}. 
For example, the two-step cascading associated with type I
second-harmonic generation (SHG) includes the generation of the second
harmonic ($\omega + \omega = 2\omega$) followed by reconstruction of the
fundamental wave through the down-conversion frequency mixing (DFM) process
($2\omega - \omega = \omega$). These two processes are governed by one
phase-matched interaction and they differ only in the direction of power
conversion.

The idea to explore more than one simultaneous nearly phase-matched
process, or {\em double-phase-matched (DPM) wave interaction}, became
attractive only recently~\cite{assanto,koynov}, 
for the purposes of all-optical 
transistors, enhanced nonlinearity-induced phase shifts, and polarization 
switching. In particular, it was shown~\cite{koynov} that multistep 
cascading  can be achieved by two second-order nonlinear cascading processes, 
SHG and  sum-frequency mixing  (SFM), and these two processes 
can also support a novel class of multi-colour parametric 
solitons~\cite{ol}, briefly discussed below.

\subsection{Multistep cascading solitons}

To introduce the simplest model of multistep cascading,  
we consider the fundamental
beam with frequency $\omega$ entering a noncentrosymmetric nonlinear medium
with a quadratic  response.  As a first step, the second-harmonic wave
with frequency $2\omega$  is generated via the SHG process.  As a second
step,  we expect the generation  of higher order harmonics due to SFM, for
example, a third harmonic ($\omega +  2\omega = 3\omega$) or even fourth
harmonic ($2\omega + 2\omega = 4\omega$)~\cite{akhmanov}.  When both such
processes are nearly phase matched,  they can lead, via down-conversion, to
a large nonlinear phase shift of  the fundamental wave~\cite{koynov}.
Additionally, the multistep cascading can support 
{\em a novel type of three-wave spatial solitary waves} 
in a diffractive $\chi^{(2)}$ nonlinear medium, 
{\em multistep cascading solitons}.

We start our analysis with the reduced amplitude equations derived in the
slowly varying  envelope approximation with the assumption of zero
absorption of all  interacting waves (see, e.g., Ref.~\cite{koynov}).
Introducing the effect of  diffraction in a slab waveguide geometry, we
obtain
\begin{equation} \label{physeqns}
 \begin{array}{l}
  {\displaystyle 
    2 i k_{1} \frac{\partial A_{1}}{\partial z} 
    + \frac{\partial^{2} A_{1}}{\partial x^{2}} 
    + \chi_{1} A_{3} A_{2}^{\ast} e^{-i\Delta k_{3}z} 
  } \\*[9pt] {\displaystyle  \qquad\qquad\qquad\qquad
    + \chi_{2}A_{2} A_{1}^{\ast}e^{-i\Delta k_{2}z} = 0, 
  } \\*[9pt] {\displaystyle 
    4 i k_{1} \frac{\partial A_{2}}{\partial z} 
    + \frac{\partial^{2} A_{2}}{\partial x^{2}} 
    + \chi_{4} A_{3} A_{1}^{\ast} e^{-i\Delta k_{3}z} 
  } \\*[9pt] {\displaystyle  \qquad\qquad\qquad\qquad
    + \chi_{5} A_{1}^{2} e^{i\Delta k_{2}z} 
    = 0, 
  } \\*[9pt] {\displaystyle 
    6 i k_{1}\frac{\partial A_{3}}{\partial z} 
    + \frac{\partial^{2} A_{3}}{\partial x^{2}} 
    + \chi_{3}A_{2}A_{1}e^{i\Delta k_{3}z} 
    = 0, 
  } \end{array}
\end{equation}
where $\chi_{1,2} = 2k_1 \sigma_{1,2}$, $\chi_3 = 6k_1 \sigma_3$, and
$\chi_{4,5} = 4k_1 \sigma_{4,5}$, and the nonlinear coupling coefficients
$\sigma_k$ are proportional to the elements of the second-order
susceptibility tensor which we assume to satisfy the following relations
(no dispersion), $\sigma_3 = 3\sigma_1$, $\sigma_2 = \sigma_5$, and
$\sigma_4 = 2\sigma_1$.

In Eqs. (\ref{physeqns}),  $A_{1}$,$A_{2}$ and $A_{3}$ are the complex
electric field envelopes of the  fundamental harmonic (FH), second harmonic
(SH), and third harmonic (TH),  respectively, $\Delta k_{2} = 2 k_1 - k_2$
is the wavevector mismatch for the  SHG process,  and $\Delta k_{3} = k_1+
k_2 - k_3$ is the wavevector mismatch  for the SFM process.  The subscripts
`1' denote the FH wave, the subscripts `2' denote the SH wave, and the
subscripts `3', the TH wave.  Following the technique earlier employed in
Refs.~\cite{ol_bur}, we look for stationary solutions of 
Eq.~(\ref{physeqns}) and introduce the normalised  
envelope $w(z,x)$, $v(z,x)$, and $u(z,x)$ according to the relations,
\begin{equation} \label{substitute}
 \begin{array}{l} 
  {\displaystyle  
    A_{1} = \frac{\sqrt{2}\beta k_{1}}{\sqrt{\chi_{2}\chi_{5}}}e^{i\beta z} w,
  } \\*[9pt] {\displaystyle 
    A_{2} = \frac{2 \beta k_{1}}{\chi_{2}}e^{2i\beta z + i\Delta k_{2} z} v, 
  } \\*[9pt] {\displaystyle 
    A_{3} = \frac{\sqrt{2\chi_{2}}\beta
    k_{1}}{\chi_{1}\sqrt{\chi_{5}}}e^{3i\beta z + i\Delta k z} u,
  } \end{array}
\end{equation}
where $\Delta k \equiv \Delta k_{2} + \Delta k_{3}$.  Renormalising the
variables as  $z \rightarrow z/\beta$ and $x \rightarrow
x/\sqrt{2\beta k_{1}}$, we finally obtain a system of coupled equations,
\begin{equation} \label{normal}
\begin{array}{l}
{\displaystyle i\frac{\partial w}{\partial z} + \frac{\partial^{2}
w}{\partial x^{2}}  - w + w^{\ast}v + v^{\ast}u = 0,} \\*[9pt]
{\displaystyle 2i \frac{\partial v}{\partial z} + \frac{\partial^{2}
v}{\partial x^{2}} - \alpha v + \frac{1}{2} w^{2} + w^{\ast}u = 0,} \\*[9pt]
{\displaystyle 3i\frac{\partial u}{\partial z} + \frac{\partial^{2}
u}{\partial x^{2}} - \alpha_{1}u + \chi vw = 0,} \\*[9pt]
\end{array}
\end{equation}
where $\alpha = 2(2\beta + \Delta k_{2})/\beta$ and $\alpha_{1} = 3(3\beta
+ \Delta k)/ \beta$ are two dimensionless parameters that characterise the
nonlinear phase matching  between the parametrically interacting waves.
Dimensionless material parameter $\chi \equiv  \chi_{1}\chi_{3}/ \chi_{2}^2
=
9 (\sigma_1/\sigma_2)^2$ depends on the type of phase matching, and it can
take different values of order of one.  For example, when both SHG and SFM
are due to quasi-phase matching (QPM), we have $\sigma_j = (2/\pi
m)(\pi/\lambda_1 n_1)  \chi^{(2)} [\omega; (4-j)\omega; - (3-j)\omega]$,
where $j =1,2$. Then, for the first-order $(m =1)$  QPM processes (see,
e.g., Ref.~\cite{pfister}), we have $\sigma_1 = \sigma_2$,  and therefore
$\chi=9$. When SFM is due to the third-order QPM process (see, e.g., Ref.
\cite{baldi}), we should take $\sigma_1 = \sigma_2/3$, and therefore
$\chi=1$. At last, when SFM is the fifth-order QPM process, we have
$\sigma_1 = \sigma_2/5$ and $\chi=9/25$.

Dimensionless equations (\ref{normal}) present a fundamental model for
three-wave  multistep cascading solitons in the absence of walk-off.
Additionally to the type I SHG solitons (see, e.g., Refs~\cite{ol_bur}),
the multistep cascading solitons involve the phase-matched SFM interaction
($\omega + 2\omega = 3\omega$) that generates a third harmonic  wave.  
Two-parameter family of localised  solutions
consists of three mutually coupled waves. It is interesting to note that,
similar to the case of nondegenerate three-wave mixing~\cite{malomed},
Eqs.~(\ref{normal}) possess an exact solution.  To find it, we make
a substitution $w = w_{0}\, {\rm sech}^{2}(\eta x)$, $v = v_{0} \,{\rm
sech}^{2}(\eta x)$ and  $u = u_{0}\, {\rm sech}^{2}(\eta x)$,  and obtain
 unknown parameters from the following algebraic  equations
\begin{equation} \label{exactsol}
w_{0}^{2} = \frac{9v_{0}}{3+4\chi v_{0}}, \;\;\; 4\chi v_{0}^{2} + 6v_{0}
=9 , \;\;\; u_{0} = \frac{2}{3} \chi w_{0}v_{0},
\end{equation}
valid for $\eta = \frac{1}{2}$ and $\alpha = \alpha_{1} = 1$. Equations
(\ref{exactsol}) have two solutions corresponding to {\em positive} and
{\em negative}  values of the amplitude ($w_0$).  This indicates a possibility of
multi-valued solutions, even within the class of exact solutions.

\mpictf{fig:tr1}{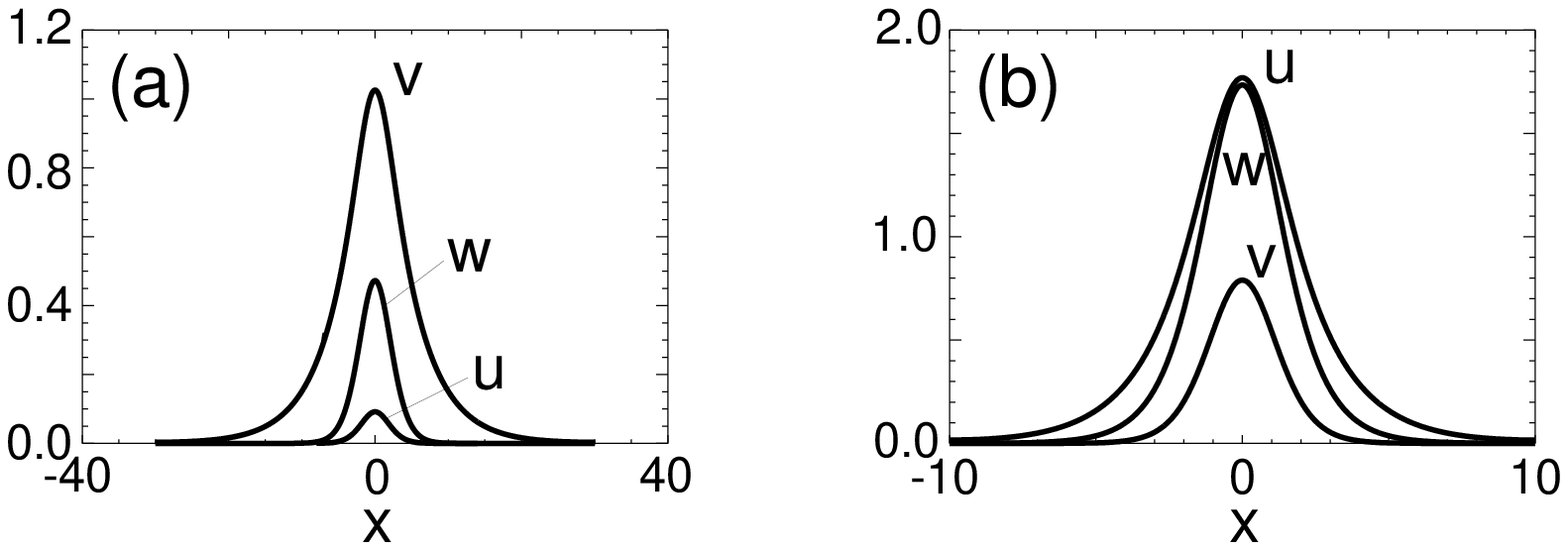}{8cm}{
Examples of three-wave solitary waves of the model (\ref{normal})
for (a) $\alpha = 0.05$,  $\alpha_1 =5$, and (b) $\alpha =5$, $\alpha_1
=0.35$.}

In general, three-wave solitons of Eqs. (\ref{normal}) can be found
only numerically.  Figures \ref{fig:tr1}(a) and \ref{fig:tr1}(b)
present two examples of solitary waves for different sets of the mismatch
parameters  $\alpha$ and $\alpha_{1}$.  When $\alpha_{1} \gg 1$ 
[see Fig.~\ref{fig:tr1}(a)], 
which corresponds to an unmatched SFM process, the amplitude of the
third harmonic is small, and it vanishes for  $\alpha_{1} \rightarrow
\infty$.

\mpictf{fig:tr2}{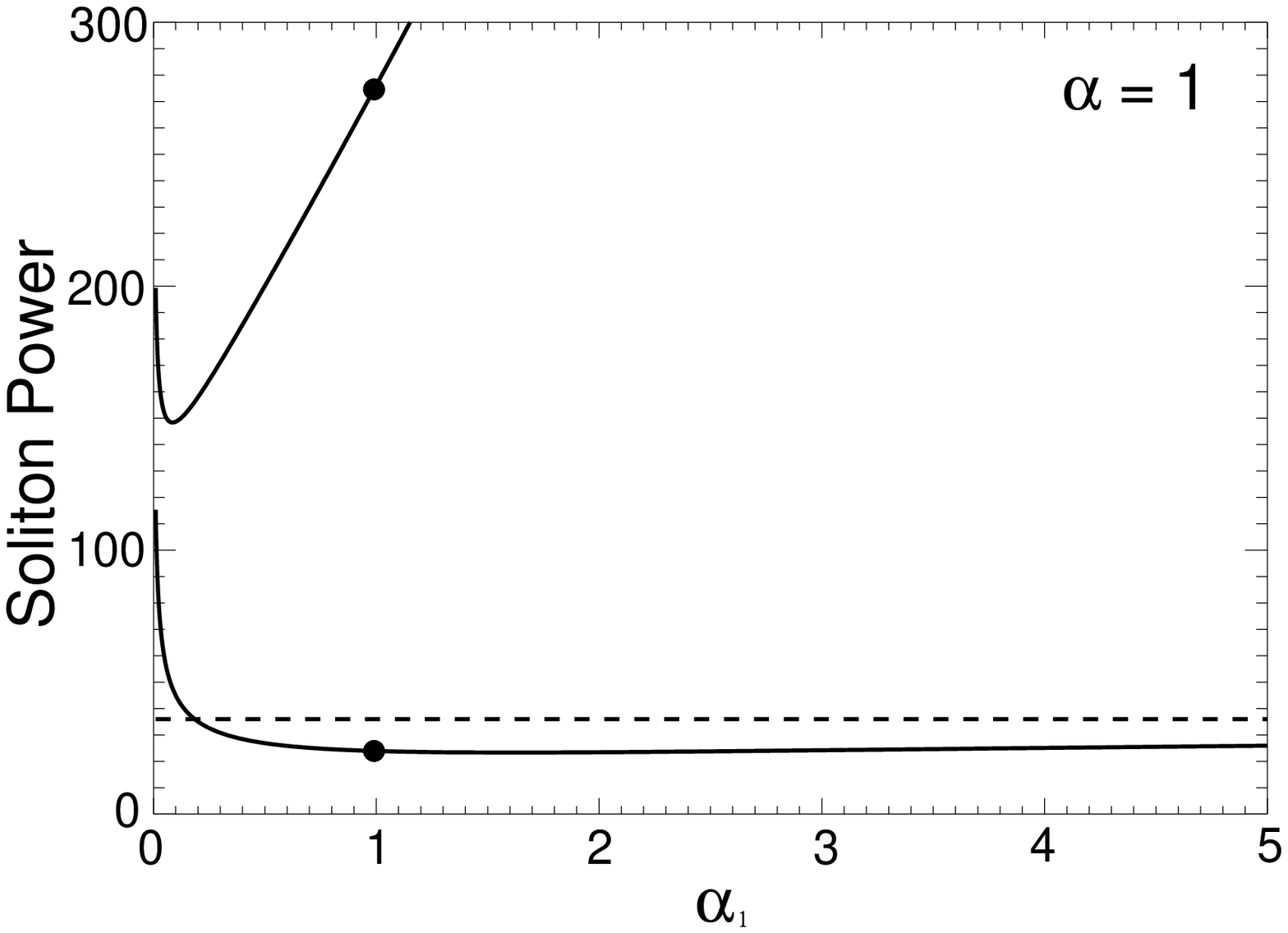}{8cm}{
Two branches of multistep cascading solitons shown as the total
soliton power $P$ vs. the parameter $\alpha_1$ for $\alpha =1$ and $\chi
=1$. Filled circles show the analytical solutions. The lower branch
approaches a family of two-wave quadratic solitons (for $\alpha_1
\rightarrow \infty$) shown by a dashed line.}

To summarise different types of three-wave solitary waves, 
in Fig.~\ref{fig:tr2} we
plot the dependence of the total soliton power defined as
\begin{equation}
\label{power}
P = \int^{+\infty}_{-\infty} dx \left( |w|^2 + 4 |v|^2 + \frac{9}{\chi}
|u|^2 \right),
\end{equation}
on the mismatch parameter $\alpha_1$, for fixed $\alpha=1$. It is clearly
seen that for some
values of $\alpha_1$ (including the exact solution at $\alpha_1 =1$ shown
by two filled circles), there exist {\em two different branches} of
three-wave solitary waves, and only one of those branches approaches, for
large values of $\alpha_1$,  a family of two-wave solitons of the cascading
limit (Fig.~\ref{fig:tr2}, dashed).  
The slope of the branches changes from negative (for
small $\alpha_1$) to positive (for large $\alpha_1$), indicating a possible
change of the soliton stability.
However, the detailed analysis of the soliton stability is beyond
the scope of this paper (see, e.g., Refs.~\cite{prl,trillo}).

\subsection{Parametric soliton-induced waveguides}

Another type of multistep cascading parametric processes which involve only 
two frequencies, i.e. {\em two-colour multistep cascading},
can occur due to the vectorial interaction of 
waves with different polarization. We denote two orthogonal polarization 
components of the fundamental harmonic (FH)
wave ($\omega_1 = \omega$) as A and B, and two orthogonal polarizations 
of the second harmonic (SH) wave ($\omega_2 = 2\omega$),
as S and T. Then, a simple multistep cascading process consists of the
following steps. First, the FH wave A generates the
SH wave S via type I SHG process. Then, by down-conversion
SA-B, the orthogonal FH wave B is generated. At last, the initial
FH wave A is reconstructed by the processes SB-A or AB-S, SA-A. 
Two principal second-order processes AA-S and AB-S correspond to {\em two
different components} of the $\chi^{(2)}$
susceptibility tensor, thus introducing additional degrees of freedom into 
the parametric interaction.
Different cases of such type of multistep cascading processes are summarized
in Table~\ref{tab:DPM}.

\noindent \parbox[l]{8cm}{
\noindent
\begin{table}
\caption{ Two-frequency multistep cascading processes}
\begin{tabular}{cll} 
   & Principal 
   & Equivalent \\ 
\hline
(a) & (AA-S, AB-S)
    & (BB-S, AB-S); (AA-T, AB-T) \\ && (BB-T, AB-T) \\
\hline
(b) & (AA-S, AB-T) 
    & (BB-S, AB-T); (AA-T, AB-S) \\ && (BB-T, AB-S) \\
\hline
(c) & (AA-S, BB-S) & (AA-T, BB-T) \\
\hline
(d) & (AA-S, AA-T) & (BB-S, BB-T) 
\end{tabular}
\label{tab:DPM}
\end{table}
}

To demonstrate some of the unique properties of the multistep cascading, we 
discuss here how it can be employed for soliton-induced waveguiding effects 
in quadratic media. 
For this purpose, we consider a model of two-frequency multistep cascading
described by the principal DPM process (c) 
(see Table~\ref{tab:DPM} above) in the planar 
slab-waveguide geometry. Using the slowly varying envelope approximation with 
the assumption of zero absorption of all interacting waves, we obtain
\begin{eqnarray} \label{eq_1}
 \begin{array}{l}
  {\displaystyle 
    2 i k_{1} \frac{\partial A}{\partial z} 
    + \frac{\partial^{2} A}{\partial x^{2}} 
    + \chi_{1} S A^{\ast} e^{-i\Delta k_1 z} 
    = 0,
  } \\*[9pt] {\displaystyle 
    2 i k_{1}\frac{\partial B}{\partial z} + \frac{\partial^{2} B}
     {\partial x^{2}} + \chi_2 S B^{\ast}e^{-i\Delta k_2 z} = 0,
  } \\*[9pt] {\displaystyle 
    4 i k_{1} \frac{\partial S}{\partial z} 
    + \frac{\partial^{2} S}{\partial x^{2}} 
    + 2 \chi_1 A^2 e^{i\Delta k_1 z} 
    + 2 \chi_2 B^2 e^{i\Delta k_2 z} 
    = 0,
  } \end{array}
\end{eqnarray}
where $\chi_{1,2} = 2k_1 \sigma_{1,2}$, the nonlinear coupling
coefficients $\sigma_k$ are proportional to the elements of the
second-order susceptibility tensor,
and $\Delta k_1$ and $\Delta k_2$ are the corresponding
wave-vector mismatch parameters.

To simplify the system (\ref{eq_1}), we look for its stationary solutions and
introduce the normalized envelopes $u$, $v$, and $w$ according to the
following relations, 
$A = \gamma_1 u \, \exp ( i\beta z - \frac{i}{2} \Delta k_1 z)$, 
$B = \gamma_2 v \, \exp (i\beta z - \frac{i}{2} \Delta k_2 z)$, and 
$S = \gamma_3 w \, \exp (2i\beta z)$, where $\gamma_1^{-1} = 2\chi_1 x_0^2$, 
$\gamma_2^{-1} = 2x_0^2 (\chi_1 \chi_2)^{1/2}$, and 
$\gamma_3^{-1} = \chi_1 x_0^2$, and the longitudinal and transverse 
coordinates are measured in the units
of $z_0 = (\beta - \Delta k_1/2)^{-1}$ and $x_0 = (z_0/2k_1)^{1/2}$,
respectively. Then, we obtain a system of normalized equations,
\begin{equation} \label{eq_N}
 \begin{array}{l}
  {\displaystyle 
   i \frac{\partial u}{\partial z} + 
    \frac{\partial^{2}u}{\partial x^{2}}  - u + u^{\ast} w = 0,} 
                     \\*[9pt]
  {\displaystyle 
   i \frac{\partial v}{\partial z} + 
    \frac{\partial^{2} v}{\partial x^{2}} - \alpha_1 v + \chi v^{\ast}w= 0,} 
                     \\*[9pt]
  {\displaystyle 
   2i \frac{\partial w}{\partial z} + 
    \frac{\partial^{2}w}{\partial x^{2}} - \alpha w+\frac{1}{2}(u^2+v^2)= 0,}
  \end{array}
\end{equation}
where $\chi = (\chi_2/\chi_1)$, 
$\alpha_1 = (\beta - \Delta k_2/2)(\beta - \Delta k_1/2)^{-1}$, and 
$\alpha = 4\beta \, (\beta - \Delta k_1/2)^{-1}$.

First of all, we notice that for $v=0$ (or, similarly, $u=0$), the 
dimensionless Eqs.~(\ref{eq_N}) reduce to the corresponding model for 
the two-step cascading due to type I SHG discussed earlier 
\cite{stegeman,chi2_review}, and its stationary solutions are defined by 
the equations for real $u$ and $w$,
\begin{equation} \label{eq_2}
 \begin{array}{l}
  {\displaystyle 
  \frac{d^2 u}{d x^2}   - u + u w  = 0,} 
          \\*[9pt]
  {\displaystyle 
  \frac{d^2 w}{d x^{2}} - \alpha w + \frac{1}{2} u^2 = 0,}
  \end{array}
\end{equation}
that possess a one-parameter family of two-wave localized solutions
$(u_0, w_0)$ found earlier numerically for any $\alpha \neq 1$, and also
known analytically for $\alpha =1$, 
$u_0(x) = {\left( 3/\sqrt{2} \right)} {\rm sech}^{2} (x/2) = \sqrt{2} w_0(x)$ 
(see Ref.~\cite{chi2_review}).

Then, in the small-amplitude approximation,
the equation for real orthogonally polarized FH wave $v$ can
be treated as an eigenvalue problem for an effective waveguide created
by the SH field $w_0(x)$,
\begin{equation} \label{eq_eigen}
  \frac{d^2 v}{d x^2} + [\chi \, w_0(x) - \alpha_1] v = 0.
\end{equation}
Therefore, an additional parametric process allows to propagate a probe beam 
of one polarization in {\em an effective waveguide} created
by a two-wave spatial soliton in a quadratic medium with FH component of 
another polarization.
However, this type of waveguide is
different from what has been studied for Kerr-like solitons because it is
{\em coupled parametrically} to the guided modes and, as a result, the 
physical picture of the guided modes is valid, rigorously speaking, only 
in the case of stationary phase-matched beams. As a result, the stability 
of the corresponding waveguide and localized modes of the orthogonal
polarization it guides is a key issue. In particular, the waveguide itself 
(i.e. {\em two-wave parametric soliton}) becomes unstable for 
$\alpha < \alpha_{\rm cr} \approx 0.2$~\cite{prl}.

In order to find the guided modes of the parametric waveguide
created by a two-wave quadratic soliton, we have to solve 
Eq.~(\ref{eq_eigen}) where the exact solution $w_0(x)$ is 
to be found numerically. 
Then, to address this problem analytically, approximate solutions can be
used, such as those found with the help of the variational 
method~\cite{variational}. However, the different types of the 
variational ansatz used do not provide a very good approximation 
for the soliton profile at all $\alpha$.
For our eigenvalue problem (\ref{eq_eigen}), the function $w_0(x)$ defines 
parameters of the guided modes and, in order to obtain accurate results, 
it should be calculated as close as possible to the exact solutions 
found numerically.
To resolve this difficulty, below we suggest a novel ``almost exact''
solution that {\em would allow to solve analytically many of the 
problems involving quadratic solitons}, including the eigenvalue 
problem~(\ref{eq_eigen}). 

First, we notice that from the exact solution at 
$\alpha =1$ and the asymptotic result for large $\alpha$, 
$w \approx u^2 / {\left( 2\alpha \right)} \approx 2 \; {\rm sech}^2 (x)$, 
it follows that the SH component 
$w_0(x)$ of Eqs. (\ref{eq_2}) remains almost self-similar for 
$\alpha \geq 1$. Thus, we look for the SH field in 
the form $w_0(x) = w_m \, {\rm sech}^2 (x/p)$, where $w_m$ and $p$ are 
to be defined. The solution for $u_0(x)$ should be consistent with 
this choice of the shape for SH, and it is defined by the 
first (linear for $u$) equation of the system (\ref{eq_2}). 
Therefore, we can take $u$ in the form of the lowest guided mode, 
$u_0(x) = u_m \, {\rm sech}^p(x/p)$, 
that corresponds to an effective waveguide $w_0(x)$. By matching the 
asymptotics of these trial functions with those defined directly  from 
Eqs.~(\ref{eq_2}) at small and large $x$, we obtain the following solution,
\begin{equation} \label{eq_S}
  u_0(x) = {u_m}{{\rm sech}^{p}(x/p)}, \;\;\; 
  w_0(x) = {w_m}{{\rm sech}^{2}(x/p)},
\end{equation}
\begin{equation} \label{eq_P}
    u_m^2 = \frac{\alpha w_m^2}{\left( w_m -1 \right)}, \;
    p= \frac{1}{\left( w_m-1 \right)}, \;
    \alpha = \frac{4 {\left( w_m-1 \right)}^3}{\left( 2-w_m \right)}.
\end{equation}
Here, the third relation allows us to find $w_m$ for 
arbitrary $\alpha$ as a solution of a cubic equation, 
and then to find all other parameters as functions of $\alpha$. 
For mismatches in the interval $0 < \alpha < +\infty$, the 
parameter values change monotonically in the regions:
$0 < u_m < +\infty$, $1 < w_m < 2$, and $+\infty > p > 1$. 
It is really amazing that the analytical solution~(\ref{eq_S}),(\ref{eq_P})  
provides {\em an excellent approximation} for the
profiles of the two-wave parametric solitons found numerically, 
with the relative errors not exceeding {1\%--3\%} for stable solitons
(e.g. when $\alpha > \alpha_{\rm cr}$).
As a matter of fact, we can treat Eqs.~(\ref{eq_S}) and~(\ref{eq_P}) 
as an {\em approximate scaling transformation} of the 
family of two-wave bright solitons. 
Moreover, this solution allows us to capture some remarkable internal 
similarities and distinctions 
between the solitons existing in different types of nonlinear media.
In particular, as follows from Eqs.~(\ref{eq_S}) and~(\ref{eq_P}), 
the FH component and the self-consistent effective waveguide (created
by the SH field) 
have approximately the same stationary transverse profiles as 
for one-component solitons in a Kerr-like medium with 
power-law nonlinear response~\cite{sulem}.
For $\alpha=1$ ($p=2$) and $\alpha \gg 1$ ($p=1$) our general
expressions reduce to the known analytical solutions, and the FH profile 
is exactly the same as that for solitons in quadratic and cubic Kerr media,
respectively. On the other hand, the strength of self-action 
for quadratic solitons depends on the normalized phase 
mismatch $\alpha$ and, in general, the beam dynamics for 
parametric wave mixing can be very different from that observed 
in Kerr-type media.

Now, the eigenvalue problem (\ref{eq_eigen}) can be readily solved 
analytically. The eigenmode cutoff values are defined by the  
parameter $\alpha_1$ that takes one of the discrete values,
$\alpha_1^{(n)} =(s-n)^2/p^2$, where 
$s= - (1/2) + [(1/4) + w_m \chi p^2]^{1/2}$.
Number $n$ stands for the mode order $(n = 0,1, \ldots)$, and the 
localized solutions are possible provided $n <s$. The profiles of the 
corresponding guided modes are
\[
   v_n(x) = V {\rm sech}^{s-n}(x/p) H(-n, 2s-n+1, s-n+1; \zeta/2),
\]
where $\zeta = 1-\tanh(x/p)$, 
$H$ is the hypergeometric function, and
$V$ is the mode's amplitude which cannot be determined within the framework of the linear analysis.

\mpictf{fig:al1}{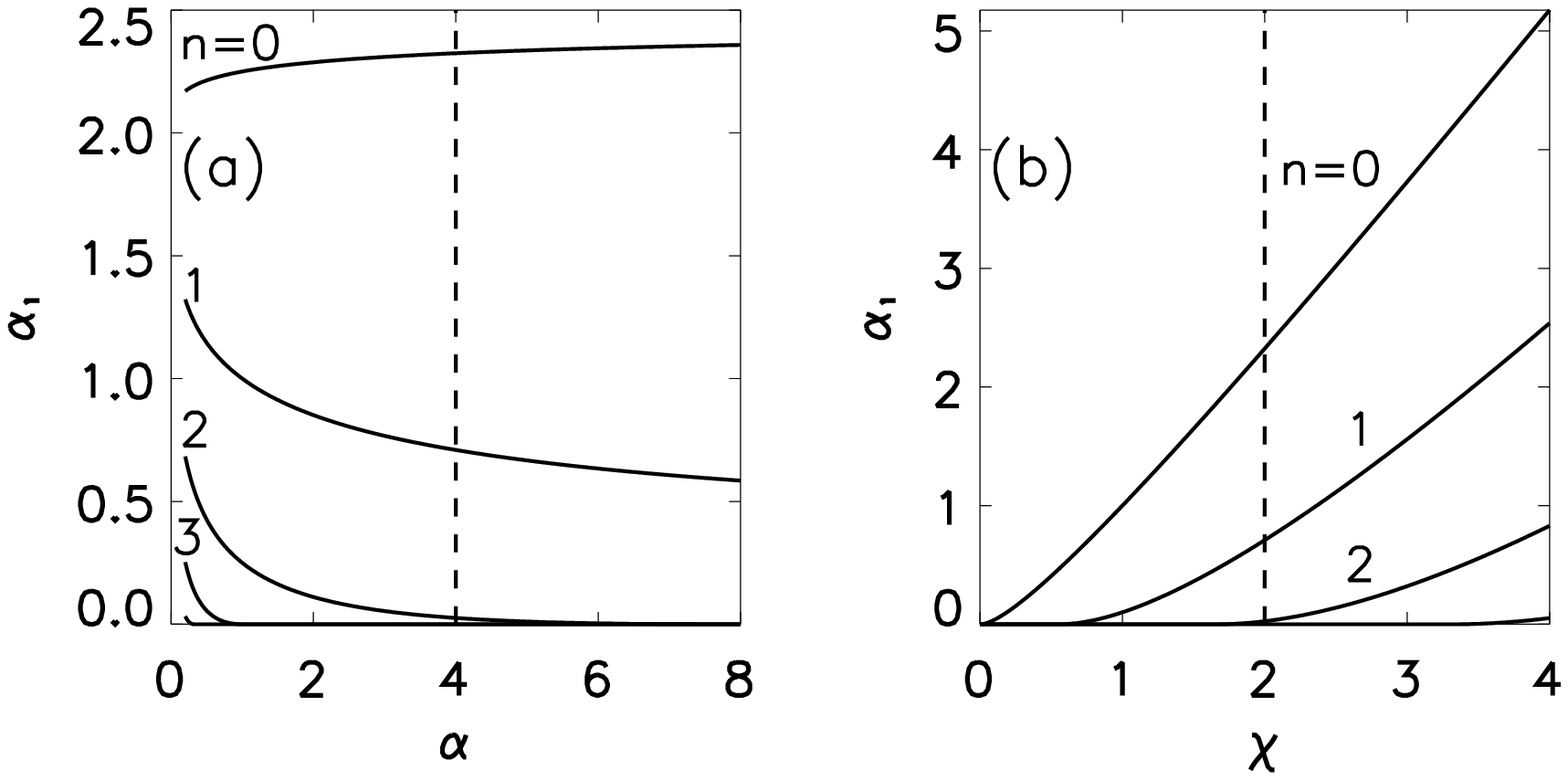}{8cm}{
Cutoff eigenvalues $\alpha_1^{(n)}$ of the guided modes shown
as (a)~functions of $\alpha$ at $\chi=2$, and (b)~functions of $\chi$ at
$\alpha=4$. Dashed lines correspond to the intersection of the plots in
the parameter space $\left( \alpha, \chi \right)$.}

According to these results, a two-wave parametric soliton creates, a multi-mode waveguide and
larger number of the guided modes can be  
observed for smaller $\alpha$. Figures~\ref{fig:al1}(a,b) show 
the dependence of the mode cutoff values $\alpha_1^{(n)}(\alpha)$  
 for a fixed $\chi$, and $\alpha_1^{(n)}(\chi)$ for a fixed $\alpha$, 
respectively.  For the case $\chi=1$, the dependence has a simple form:
$\alpha_1^{(n)}(\alpha) = [1-n(w_m-1)]^2$.

Because a two-wave soliton creates an induced waveguide parametrically 
coupled to its guided modes of the orthogonal polarization, the dynamics 
of the guided  modes {\em may differ drastically} from that of conventional 
waveguides based on the Kerr-type nonlinearities. 
Figures~\mbox{\ref{fig:wave_w}(a-d)}  
show two examples of the evolution of guided modes. 
In the first example [see Fig.~\ref{fig:wave_w}(a-c)], 
a weak fundamental mode is amplified via parametric interaction 
with a soliton waveguide, and the mode experiences a strong power 
exchange with the orthogonally polarized FH component through the SH field. This process is accompanied by only a weak deformation of the induced waveguide
[see Fig.~\ref{fig:wave_w}(a) -- dotted curve]. The resulting effect 
can be interpreted as a power exchange between two guided modes of
orthogonal polarizations in a waveguide created by the SH field.
In the second example, the propagation is stable 
[see Fig.~\ref{fig:wave_w}(d)]. 

When all the fields in Eq.~(\ref{eq_N}) are not small, i.e. the
small-amplitude approximation is no longer valid,
the profiles of the three-component solitons should be found 
numerically.
However, some of the lowest-order states can be calculated
approximately using the approach of the ``almost exact'' 
solution (\ref{eq_S}),(\ref{eq_P}) described above, which is presented
in detail elsewhere~\cite{andrey,our_pre}.
Moreover, a number of the solutions and their families can be 
obtained in {\em an explicit analytical form}. For example, for 
$\alpha_1=1/4$, there exist two
{\em families of three-component solitary waves}
for any $\alpha \geq 1$, that describe
soliton branches starting at the bifurcation points $\alpha_1=\alpha_1^{(1)}$
at \mbox{$\alpha=1$}: (i)~the soliton with a zero-order guided mode 
for $\chi=1/3$:
$u(x) = {\left( 3/\sqrt{2} \right)}\, {\rm sech}^2\left( x/2 \right)$,
$v(x) = c_2\, {\rm sech}\left( x/2 \right)$,
$w(x) = \left( 3/2 \right)\, {\rm sech}^2\left( x/2 \right)$,
and (ii)~the soliton with a first-order guided mode 
for $\chi=1$:
$u(x) = c_1\, {\rm sech}^2\left( x/2 \right)$,
$v(x) = c_2\, {\rm sech}^2\left( x/2 \right) {\rm sinh}\left( x/2 \right)$,
$w(x) = \left( 3/2 \right)\, {\rm sech}^2\left( x/2 \right)$,
where 
$c_2 = \sqrt{3 \left( \alpha-1 \right)}$ and
$c_1 = \sqrt{\left( 9/2 \right) + c_2^2}$.

\mpictf{fig:wave_w}{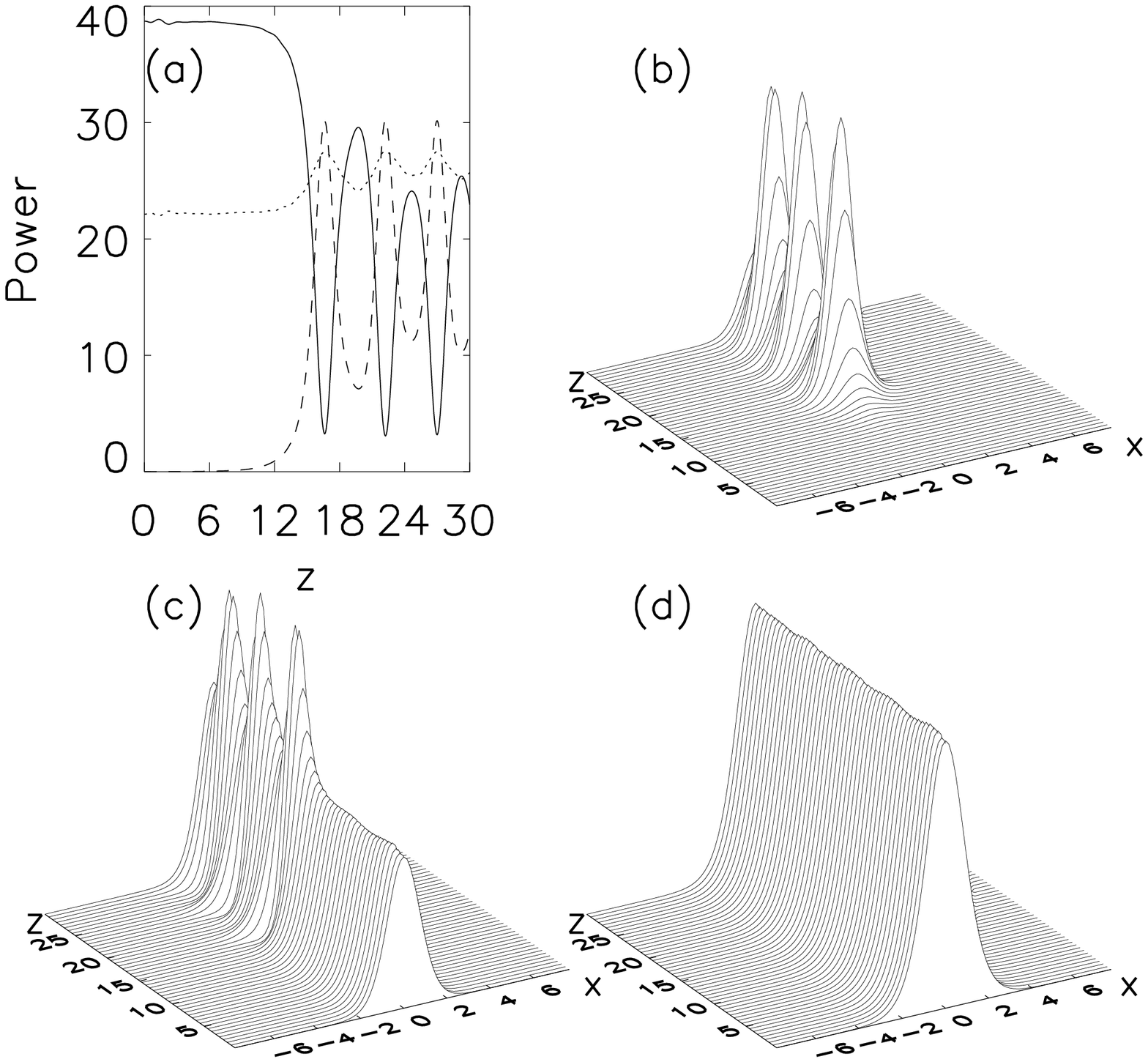}{8cm}{
(a)~Change of the normalized power in FH ($u$, solid) and 
SH ($w$, dotted) components, which initially constitute a two-wave 
soliton,
and in the guided mode ($v$, dashed) at $\chi=2$, demonstrating 
amplification of a guided wave.
Evolution of 
the guided wave and effective 
waveguide (SH) is presented in plots (b) and (c), respectively.
(d)~Stationary propagation of a stable fundamental mode ($\chi=1$).
For all the plots $\alpha = 4$, the initial amplitude 
is $v_0 = 0.1$, and $\alpha_1$ corresponds to the bifurcation point. }

For a practical realization of the DPM processes and the soliton
waveguiding effects described above, 
we can suggest two general methods. The first method 
is based on the use of {\em  two commensurable periods} of the 
quasi-phase-matched (QPM) periodic grating.  Indeed, to achieve DPM, we 
can  employ the first-order QPM for one parametric process, and the 
third-order QPM,  for the other parametric process. Taking, as an example, 
the parameters for LiNbO$_3$ and AA-S $(xx-z)$ and BB-S $(zz-z)$ processes
\cite{book_dmitr}, we find two points for DPM at about 0.89 $\mu$m and 1.25 $\mu$m. 
This means that a  single QPM grating can provide simultaneous phase-matching 
for two parametric  processes. For such a configuration, 
we obtain  $\chi \approx 1.92$ or, interchanging the polarization 
components, $\chi \approx 0.52$.
The second method to achieve the conditions of DPM processes is based on the 
idea of  {\em quasi-periodic QPM grating}~\cite{Fib_exp,THG}.
Specifically, Fibonacci optical 
superlattices provide an effective way to achieve phase-matching 
at {\em several incommensurable periods} allowing multi-frequency harmonic 
generation in a single structure. We describe the properties
of such structures in the next section.

\section{ENVELOPE SOLITONS IN FIBONACCI SUPERLATTICES}
\subsection{Incoherence and solitary waves}

For many years, solitary waves have been 
considered as {\em coherent localized modes} of nonlinear systems,
with particle-like dynamics quite dissimilar to the irregular and
stochastic behavior observed for chaotic systems~\cite{book}. 
However, about 20 years ago Akira Hasegawa, while developing a 
statistical description of the dynamics of an ensemble of plane 
waves in nonlinear strongly dispersive plasmas, suggested the 
concept of a localized envelope of random phase waves~\cite{hasegawa}. 
Because of the relatively high powers required for generating 
self-localized random waves, this notion remained a theoretical 
curiosity until recently, when the possibility to 
generate spatial optical solitons by a partially incoherent source 
was discovered in a photorefractive medium~\cite{inc_exp}.

The concept of incoherent solitons can be compared with a different 
problem: the propagation of a soliton through a spatially 
disordered medium. 
Indeed, due to random scattering on defects, the phases of the 
individual components forming a soliton experience random 
fluctuations, and the soliton itself becomes {\em partially 
incoherent} in space and time.
For a low-amplitude wave (linear regime) spatial incoherence is
known to lead to a fast decay. 
As a result, the transmission coefficient vanishes exponentially 
with the length of the system, the phenomenon known as Anderson 
localization~\cite{gredeskul}. 
However, for large amplitudes (nonlinear regime), when the 
nonlinearity length is much smaller than the Anderson localization 
length, a soliton can propagate almost unchanged through a 
disordered medium as predicted theoretically in 1990 
\cite{my_prl} and recently verified experimentally~\cite{hopkins}.

These two important physical concepts, spatial self-trapping of light 
generated by an incoherent source in a homogeneous medium, and 
suppression of Anderson localization for large-amplitude waves in 
spatially disordered media, both result from the effect of 
strong nonlinearity.
When the nonlinearity is sufficiently strong it acts as {\em an 
effective
phase-locking mechanism} by producing a large frequency shift of the
different random-phase components, and thereby introducing {\em an 
effective order} into an incoherent wave packet, thus enabling the 
formation of localized structures. In other words, both phenomena 
correspond to the limit when the ratio of 
the nonlinearity length to the characteristic length of (spatial or 
temporal) fluctuations is small. 
In the opposite limit, when this ratio is large, the wave 
propagation is practically linear.

Below we show that, at least for aperiodic inhomogeneous 
structures, solitary waves can exist in the intermediate regime 
in the form of {\em quasiperiodic nonlinear localized modes}. As an example,
we consider SHG and nonlinear beam propagation 
in {\em Fibonacci optical superlattices}, and demonstrate 
numerically the possibility of spatial self-trapping of quasiperiodic 
waves whose envelope amplitude varies quasiperiodically, while still 
maintaining a stable, well-defined spatially localized structure, 
{\em a quasiperiodic envelope soliton}.

\subsection{Quasi-phase-matching in optical superlattices}

We consider the interaction of a fundamental wave with the 
frequency $\omega$ (FH) and its SH in a slab 
waveguide with quadratic (or $\chi^{(2)}$) nonlinearity. 
Assuming the $\chi^{(2)}$ susceptibility to be modulated and the 
nonlinearity to be of the same order as diffraction, we write 
the dynamical equations in the form
\begin{equation} \label{dynam}
  \begin{array}{l} 
  {\displaystyle i\frac{\partial u}{\partial z} + \frac{1}{2} 
  \frac{\partial^2 u}{\partial x^2} + d(z) u^{\ast} v e^{-i\beta z} 
  =  0,} \\*[9pt] 
  {\displaystyle i\frac{\partial w}{\partial z} + \frac{1}{4} 
  \frac{\partial^2 w}{\partial x^2} + d(z) u^2 e^{i\beta z} =  0,}
  \end{array}
\end{equation}
where $u(x,z)$ and $w(x,z)$ are the slowly varying envelopes of the
FH and SH, respectively.
The parameter $\beta= \Delta k |k_{\omega}|x_0^2$ is proportional to 
the phase mismatch $\Delta k= 2k_{\omega}-k_{2\omega}$, $k_{\omega}$ 
and $k_{2\omega}$ being the wave numbers at the two frequencies. 
The transverse coordinate $x$ is measured in units of the input beam 
width $x_0$, and the propagation distance $z$ in units of the 
diffraction length $l_d= x_0^2|k_{\omega}|$. 
The spatial modulation of the $\chi^{(2)}$ susceptibility 
is described by the quasi-phase-matching (QPM) grating function 
$d(z)$. 
In the context of SHG, the QPM technique is an effective way to achieve 
phase matching, and it has been studied intensively~\cite{QPM}.

\mpictf{fig:d_z}{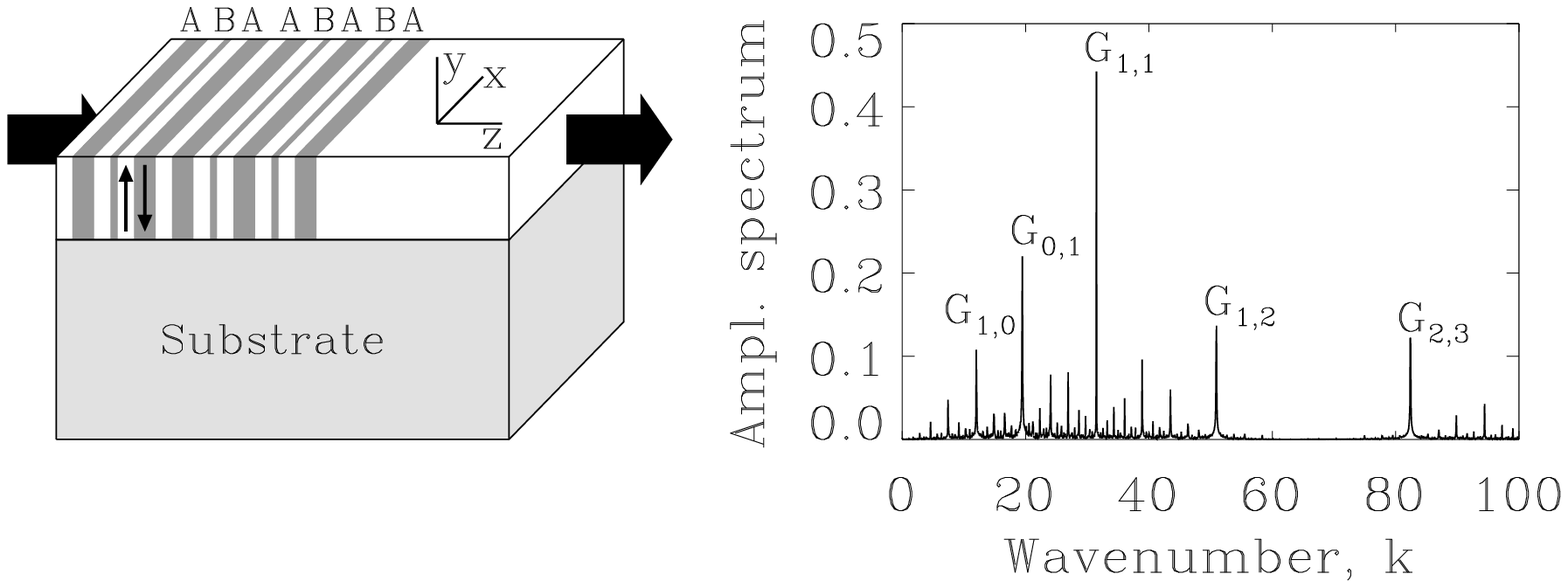}{8cm}{
(a) Slab waveguide with quasiperiodic QPM superlattice structure 
composed of building blocks A and B.
(b) Numerically calculated amplitude spectrum of $d(z)$.}

Here we consider a QPM grating produced by a quasiperiodic nonlinear
optical superlattice. 
Quasiperiodic optical superlattices, one-dimensional analogs of 
quasicrystals~\cite{quasi}, are usually designed to study the effect
of Anderson localization in the linear regime of light propagation. 
For example, Gellermann {\em et al.} measured the optical transmission 
properties of quasiperiodic dielectric multilayer stacks of SiO$_2$ 
and TiO$_2$ thin films and observed a strong suppression of the
transmission~\cite{geller}. 
For QPM gratings, a nonlinear quasiperiodic superlattice of
LiTaO$_3$, in which two antiparallel ferro-electric domains are 
arranged in a Fibonacci sequence, was recently fabricated by Zhu 
{\em et al.}~\cite{Fib_exp}, who measured multi-colour SHG with 
energy conversion efficiencies of $\sim 5\%-20\%$.
This quasiperiodic optical superlattice in LiTaO$_3$ can 
also be used for efficient direct third harmonic generation
\cite{THG}.

The quasiperiodic QPM gratings have two building blocks A and B 
of the length $l_A$ and $l_B$, respectively, which are ordered 
in a Fibonacci sequence [Fig.~\ref{fig:d_z}(a)]. 
Each block has a domain of length $l_{A_1}$=l ($l_{B_1}$=l) with
$d$=$+1$ (shaded) and a domain of length $l_{A_2}$=$l(1+\eta)$ 
[$l_{B_2}$=$l(1-\tau\eta)$] with $d$=$-1$ (white).
In the case of $\chi^{(2)}$ nonlinear QPM superlattices this 
corresponds to positive and negative ferro-electric domains, 
respectively.
The specific details of this type of Fibonacci optical superlattices 
can be found elsewhere~\cite{Fib_exp}.
For our simulations presented below we have chosen $\eta$= $2(\tau-1)
/(1+\tau^2)$= 0.34, where $\tau$= $(1+\sqrt{5})/2$ is the so-called 
{\em golden ratio}. 
This means that the ratio of length scales is also the golden ratio, 
$l_A/l_B$= $\tau$. 
Furthermore, we have chosen $l$=0.1.

The grating function $d(z)$, which varies between $+1$ and $-1$
according to the Fibonacci sequence, can be expanded in a Fourier 
series
\begin{equation}
   \hspace{-7mm}
   d(z)= \sum_{m,n}d_{m,n}e^{iG_{m,n}z},\;\;\;
   G_{m,n}= \frac{2\pi(m+n\tau)}{D},
   \label{Gmn}
\end{equation}
where $D$=$\tau l_A+l_B$=0.52 for the chosen parameter values.
Hence the spectrum is composed of sums and differences of
the basic wavenumbers $\kappa_1$=$2\pi/D$ and $\kappa_2$=$2\pi\tau/D$.
These components fill the whole Fourier space densely, since $\kappa_1$
and $\kappa_2$ are incommensurate. Figure~\ref{fig:d_z}(b) shows the
numerically calculated Fourier spectrum $G_{m,n}$. 
The lowest-order ``Fibonacci modes'' are clearly the most intense. 

\subsection{Quasiperiodic optical solitons}

To analyze the beam propagation and SHG in a quasiperiodic QPM 
grating one could simply average Eqs.~(\ref{dynam}).
To lowest order this approach always yields a system of
equations with constant mean-value coefficients, which does not 
allow to describe oscillations of the beam amplitude and phase. 
However, here we wish to go beyond the averaged equations and consider
the rapid large-amplitude variations of the envelope functions. 
This can be done analytically for periodic QPM gratings~\cite{qpm}.
However, for the quasiperiodic gratings we have to resolve 
to numerical simulations.

Thus we have solved Eqs.~(\ref{dynam}) numerically with a second-order
split-step routine.
At the input of the crystal we excite the fundamental beam
(corresponding to unseeded SHG) with a Gaussian profile, 
\begin{equation}
   u(x,0) =  A_u \, e^{-x^2/10}, \;\;\; w(x,0) =  0.
   \label{initial}
\end{equation}
We consider the quasiperiodic QPM grating with matching to the 
peak at $G_{2,3}$, i.e., $\beta$=$G_{2,3}$=82.25. 
First, we study the small-amplitude limit when a weak FH is injected 
with a low amplitude. 
Figures~\ref{fig:soliton}(a,b) show an example of the evolution of FH 
and SH in this effectively linear regime.
As is clearly seem from Fig.~\ref{fig:soliton}(b) the SH wave is excited, 
but both beams eventually diffract.

When the amplitude of the input beam exceeds a certain threshold,
self-focusing and localization should be observed for both harmonics.
Figures~\ref{fig:soliton}(c,d) show an example of the evolution of a strong
input FH beam, and its corresponding SH. 
Again the SH is generated, but now the nonlinearity is so
strong that it leads to self-focusing and mutual self-trapping of the 
two fields, resulting in a spatially localized two-component soliton,
despite the continuous scattering of the quasiperiodic QPM 
grating.

\mpictf{fig:soliton}{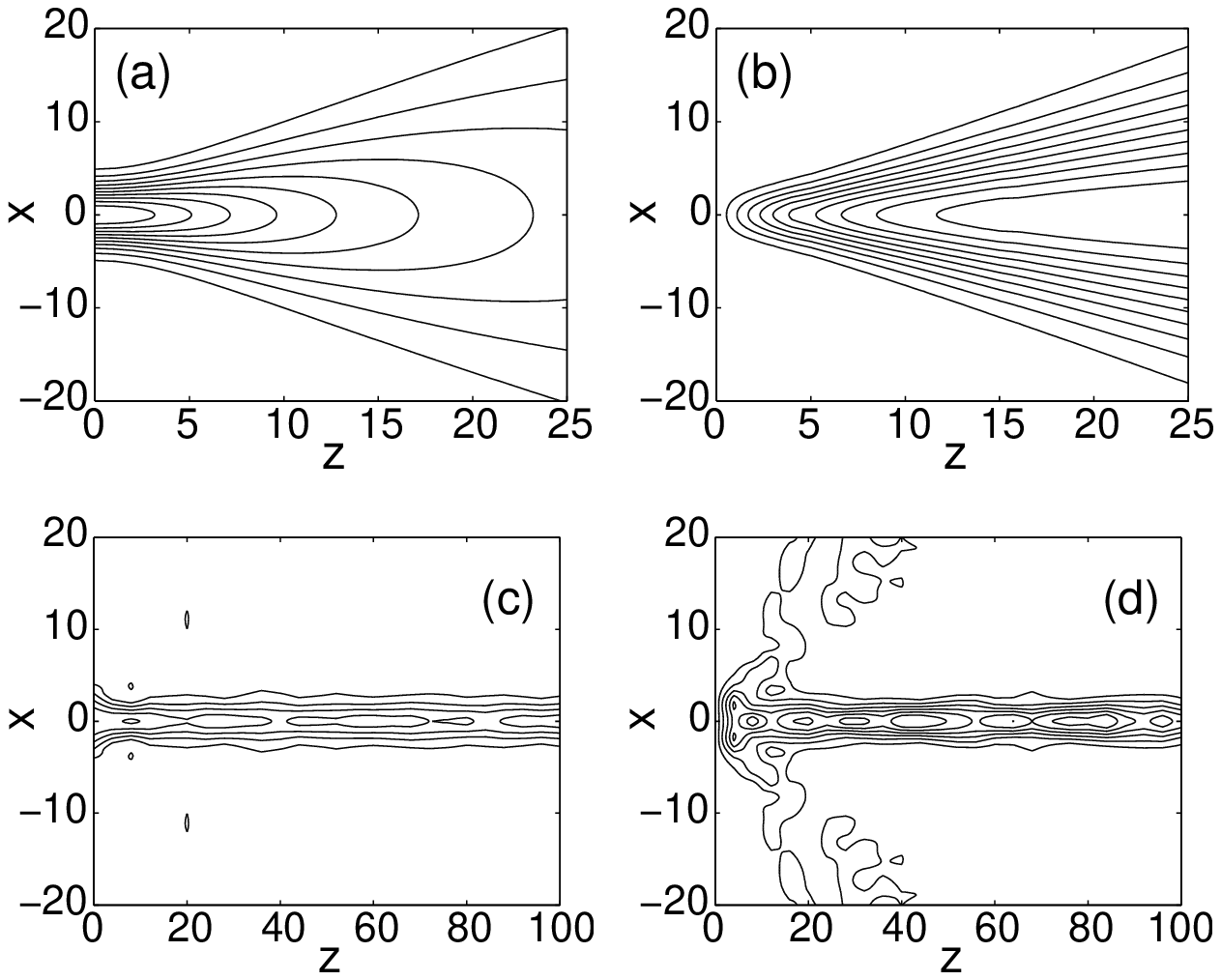}{8cm}{
(a) Diffraction of a weak FH beam with amplitude $A_u$=0.25 
for $\beta$=82.25. (c) Excitation of a quasiperiodic soliton by a FH 
beam with amplitude $A_u$=5 for $\beta$=82.25. (b,d) Corresponding 
SH components.}

It is important to notice that the two-component localized beam 
created due to the self-trapping effect is quasiperiodic by itself. 
As a matter of fact, after an initial transient its amplitude
oscillates in phase with the quasiperiodic QPM modulation $d(z)$. 
This is illustrated in Fig.~\ref{fig:oscillations}, where we show 
in more detail the peak intensities in the
asymptotic regime of the evolution.
The oscillations shown in Fig.~\ref{fig:oscillations} are in phase
with the oscillations of the QPM grating $d(z)$, and we indeed found that 
their spectra are similar. 

\mpictf{fig:oscillations}{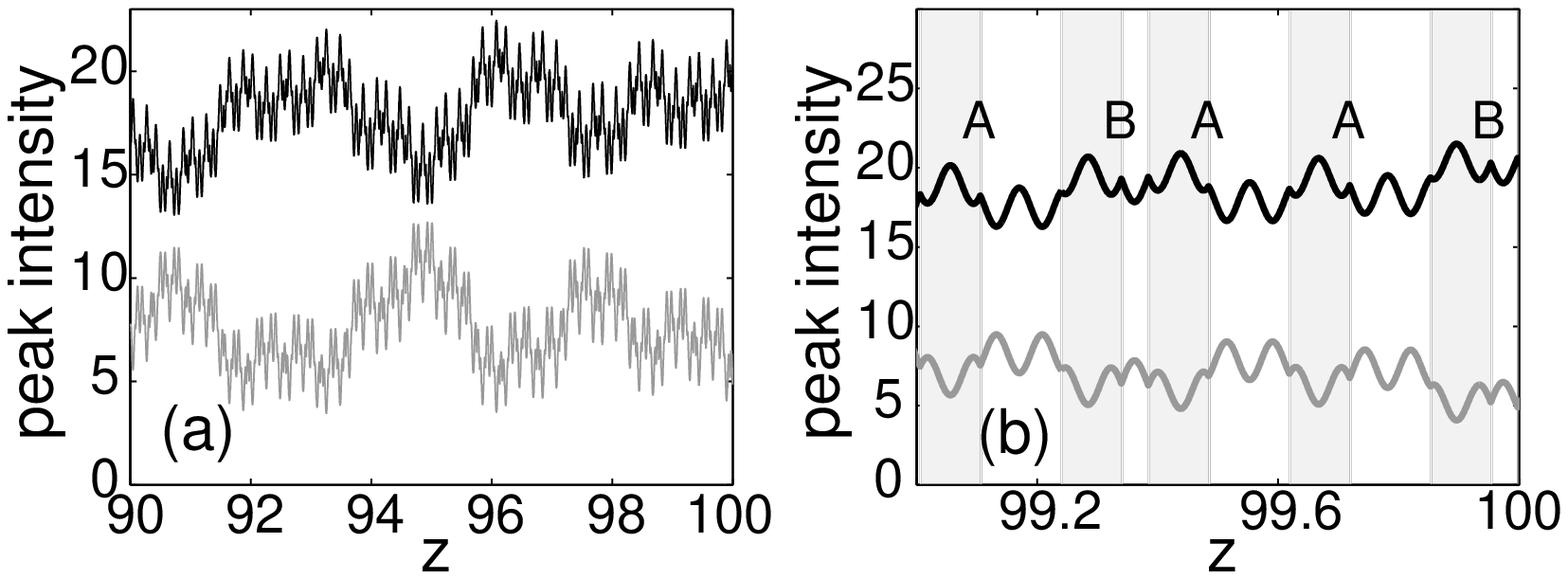}{8cm}{
Amplitude oscillations of the quasiperiodic soliton.
(a),(b) Close-ups of the peak intensity $|u(0,z)|^2$ of the FH 
(black) and $|w(0,z)|^2$ of the SH (grey). 
The Fibonacci building blocks A and B are indicated in (b) with 
$d$=1 in grey regions, and $d$=$-1$ in white regions.}

Our numerical results show that the quasiperiodic envelope solitons 
can be generated for a broad range of the phase-mismatch $\beta$.
The amplitude and width of the solitons depend on the effective
mismatch, which is the separation between $\beta$ and the nearest
strong peak $G_{m,n}$ in the Fibonacci QPM grating spectrum
[see Fig.~\ref{fig:d_z}(b)].
Thus, low-amplitude broad solitons are excited for $\beta$-values
in between peaks, whereas high-amplitude narrow solitons are
excited when $\beta$ is close to a strong peak, as shown in
Fig.~\ref{fig:soliton}(c,d).

\mpictf{fig:transmission}{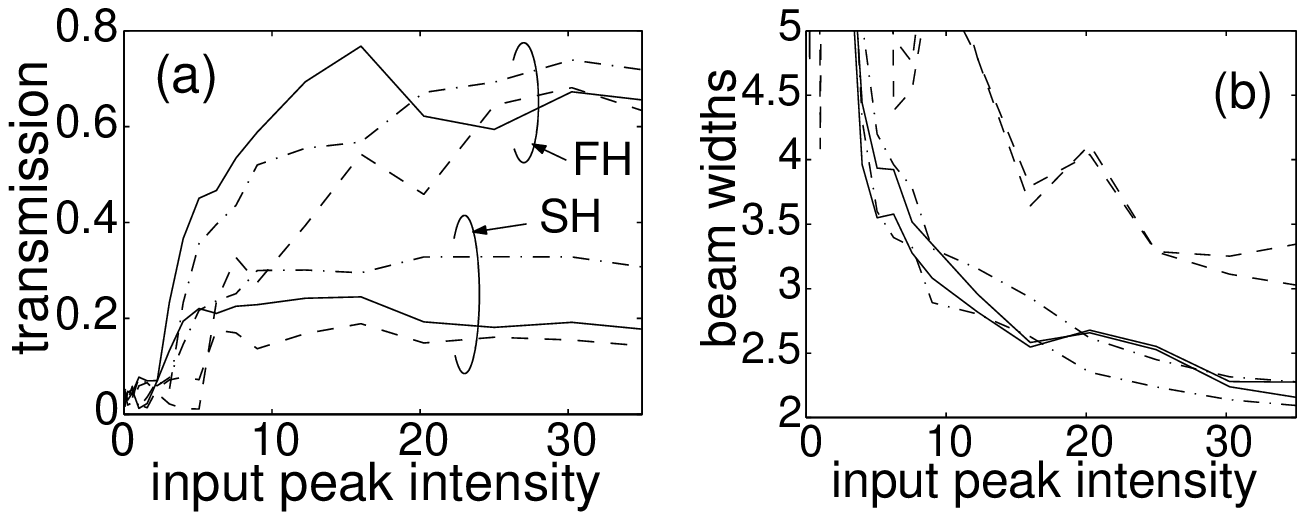}{8cm}{
(a) Transmission of the FH, $|u(0,L)/u(0,0)|^2$, and the SH,
$|w(0,L)/u(0,0)|^2$ vs.\ input peak intensity $|u(0,0)|^2$ 
of the FH. (b) Output beam width vs.\ $|u(0,0)|^2$. $L$=100, 
$\beta$=$G_{1,2}$=50.83 (solid), $\beta$=$G_{2,4}$=101.66
(dashed), and $\beta$=$G_{2,3}$=82.25 (dot-dashed).}

To analyse in more detail the transition between the linear 
(diffraction) and nonlinear (self-trapping) regimes, we have made
a series of careful numerical simulations~\cite{cbc}. 
In Fig.~\ref{fig:transmission} we show the transmission coefficients and 
the beam widths at the output of the crystal versus the intensity 
of the FH input beam, for a variety of $\beta$-values.
These dependencies clearly illustrate the 
universality of the generation of localised modes for varying strength 
of nonlinearity, i.e. a quasiperiodic
soliton is generated only for sufficiently high amplitudes. 
This is of course a general phenomenon also observed in many nonlinear 
isotropic media. 
However, here the self-trapping occurs for quasiperiodic waves, with
the quasiperiodicity being preserved in the variation of the amplitude 
of both components of the soliton.

\section{CONCLUSION}

We have overviewed several important physical examples of the multi-component 
solitary waves which appear due to multi-mode and/or multi-frequency 
coupling in nonlinear optical fibers and waveguides.   
We have described several types of such multi-component solitary waves, 
including: 
(i)~multi-wavelength solitary waves in multi-channel 
bit-parallel-wavelength fiber transmission systems, 
(ii)~multi-colour parametric spatial solitary waves due to multistep 
cascading in quadratic materials, and 
(iii)~quasiperiodic envelope solitons in Fibonacci optical superlattices. 
These examples reveal some general features and properties of multi-component 
solitary waves in nonintegrable nonlinear models, also serving as a stepping stone for approaching other problems of the multi-mode soliton coupling and interaction. 

\section*{ACKNOWLEDGMENTS}       
 
The work was supported by the Australian Photonics Cooperative Research 
Centre and by a collaborative Australia-Denmark grant of the Department of 
Industry, Science, and Tourism (Australia).

\end{multicols}
\end{document}